# Genetic Algorithm Based Optimization of Clustering in Ad-Hoc Networks

[*1]Bhaskar Nandi     [#2]Subhabrata Barman     [#3]Soumen Paul

[#2,#3]Haldia Institute of Technology, W.B, India.

#2 ____________________
#3 ____________________
*1 ____________________

*Abstract*:

**In this paper, we have to concentrate on implementation of Weighted Clustering Algorithm with the help of Genetic Algorithm (GA).Here we have developed new algorithm for the implementation of GA-based approach with the help of Weighted Clustering Algorithm (WCA) [4]. Cluster-Head chosen is a important thing for clustering in ad-hoc networks. So, we have shown the optimization technique for the minimization of Cluster-Heads(CH) based on some parameter such as degree-difference , Battery power ($P_v$), degree of mobility, and sum of the distances of a node in ad-hoc networks. Cluster-Heads selection of ad-hoc networks is an important thing for clustering. Here, we have discussed the performance comparison between deterministic approach and GA-based approach. In this performance comparison, we have seen that GA does not always give the good result compare to deterministic WCA algorithm. Here we have seen connectivity (connectivity can be measured by the probability that a node is reachable to any other node.) is better than the deterministic WCA algorithm [4].**

*Keywords*- **Adhoc Networks, GA, Cluster Head (CH), WCA.**

## I. INTRODUCTION

A wireless ad-hoc network consists of nodes that move freely and communicate with each other using wireless links. Ad-hoc networks do not use specialized routers for path discovery and traffic routing. One way to support efficient communication between nodes is to develop wireless backbone architecture; this means that certain nodes must be selected to form the backbone. Over time, the backbone must change to reflect the changes in the network topology as nodes move around. The algorithm that selects the members of the backbone should naturally be fast, but also should require as little communication between nodes as possible, since mobile nodes are often powered by batteries. One way to solve this problem is to group the nodes into clusters, where one node in each cluster functions as cluster head, responsible for routing. A clusterhead does the resource allocation to all the nodes belonging to its cluster. Due to the dynamic nature of the mobile nodes, their association and dissociation to and from clusters perturb the stability of the network and thus reconfiguration of cluster heads is unavoidable. Thus, it is desirable to have a minimum number of clusterheads that can serve the network nodes scattered evenly in the area. An optimal selection of the clusterheads is an NP-hard problem . Therefore, various heuristics have been designed for this problem . we apply genetic algorithms (GA) as an optimization technique to improve the performance of clusterhead election procedure. In particular, GAs are defined as search algorithms that use the mechanics of natural selection and genetics such as reproduction, gene crossover, mutation as their problem-solving method. The goal is to he able to find out a better solution in the form of new generations that have received advantages and survival-enhancing traits from the previous Generations. We have to target artificial-life simulation is created where survival of the fittest logic is applied for the string structures that are the living organism equivalent in real world. Even though the representation is structured, there is a randomization in data exchange to simulate the evaluation of real life forms. As each generation brings up a new set of strings by different combination of bits of pieces of the previous generation, the results are not guaranteed to come up with a generation that has a better fitness value hut by performing different genetic operations, the probability of achieving the desired results is increased.

## II. CLUSTERING IN ADHOC NETWORKS

The weight-based distributed clustering algorithm that takes into consideration that the number of nodes that a cluster head can handle the ideal degree, transmission power, mobility and battery power of a mobile node. We try to keep the number of nodes in a cluster around a pre-defined threshold to facilitate the optimal operation of the medium access control (MAC) protocol. Our cluster head election procedure is n periodic as in earlier research, but adapts based on the dynamism of threshold value of nodes. This on-demand execution of WCA aims to maintain the stability of the network, thus lowering the computation and communication cost associated with it.

A cluster head may not be able handle a large number of nodes due to resource limitations even if these nodes are its neighbors and lie well within its transmission range. Thus, the load handling capacity of the cluster head puts an upper bound on the node-degree. In other words, simply covering the area with the minimum number of cluster heads will put more burden on the cluster heads. At the same time, more cluster heads will lead to a computationally expensive system. This may result in good throughput, but the data packets have to go through multiple hops resulting in high latency. In summary, choosing an optimal number of cluster heads which will yield high throughput but incur as low latency as possible, is still an important problem. As the search for better heuristics for this problem continues, we propose the use of a combined weight metric, that takes into account several system parameters like the ideal node-degree, transmission power, mobility and the battery power of the nodes. We could have a fully distributed system where all the nodes share the same responsibility and act as cluster heads. However, more cluster heads result in extra number of hops for a packet when it gets routed from the source to the destination, since the packet has to go via larger number of cluster heads. Thus this solution leads to higher latency, more power consumption and more information processing per node. On the other hand, to maximize the resource utilization, we can choose to have the minimum number of cluster heads to cover the whole geographical area over which the nodes are distributed. The whole area can be split up into zones, the size of which can be determined by the transmission range of the nodes. This can put a lower bound on the number of cluster heads required. Ideally, to reach this lower bound, a uniform distribution of the nodes is necessary over the entire area. Also, the total number of nodes per unit area should be restricted so that the cluster head in a zone can handle all the nodes therein. However, the zone based clustering is not a viable solution due to the following reasons. The cluster heads would typically be centrally located in the zone, and if they move, new cluster heads have to be selected. It might so happen that none of the other nodes in that zone are centrally located. Therefore, to find a new node which can act as a cluster head with the other nodes within its transmission range might be difficult. Another problem arises due to non-uniform distribution of the nodes over the whole area. If a certain zone becomes densely populated then the cluster head might not be able to handle all the traffic generated by the nodes because there is an inherent limitation on the number of nodes a cluster head can handle. We propose to select the minimum number of cluster heads which can support all the nodes in the system satisfying the above constraints.

### III. CLUSTER HEAD ELECTION PROCEDURE

The network formed by the nodes and the links can be represented by an undirected graph $G=(V,E)$ where $V$ represents the set of nodes $v_i$ and $E$ represents the set of links $e_i$. Dominant set S is subset of V(G).such that

Union of $N(V)=V(G)$

Here $N(V)$ is the *neighborhood* of node $v$, defined as

$$d_v = |N(v)| = \sum_{v' \in V, v' \neq v} \{dist(v,v') < tx_{range}\}$$

where *tx range* is the transmission range of $v$.
Clustering Algorithm use a *combined weight* metric to search dominant set, the combined weight is composed by cluster head degree, battery power, mobility, distance. The *Cluster head election procedure* consists of eight steps as described below:

Step 1. Find the neighbors of each node $v$ which defines its *degree*——$dv$ as

$$d_v = |N(v)| = \sum_{v' \in V, v' \neq v} \{dist(v,v') < tx_{range}\}$$

Step2: Compute the *degree-difference* $\Delta_v = |d_v - \delta|$ for every node $v$. Here δ is ideal node number of a cluster except the cluster head.

Step3: For every node, compute the *sum of the distances*, $D v$, with all its neighbors, as

$$D_v = \sum_{v' \in N(v)} \{dist(v,v')\}$$

Step 4. Compute the running average of the speed for every node till current time $T$. This gives a measure of mobility and is denoted by $M v$, as

$$M_v = \frac{1}{T}\sum_{t=1}^{T}\sqrt{(X_t - X_{t-1})^2 + (Y_t - Y_{t-1})^2}$$

Where $(X_t, Y_t)$ and $(X_{t-1}, Y_{t-1})$ are the coordinates of the node $v$ at time $t$ and t-1 respectively.

Step 5. Compute the cumulative time, $P v$ during which a node $v$ acts as a cluster head. $P v$ implies how much battery power has been consumed which is assumed more for a cluster head than an ordinary node.

Step 6. Calculate the *combined weight* $W v$ for each node $v$,

$$W_v = w_1\Delta v + w_2 Dv + w_3 Mv + w_4 Pv$$

w1, w2, w3, w4 are the *weighing factors* for the corresponding system parameters and
    w1+ w2+ w3+ w4=1.

Step 7. Choose that node with the smallest $W v$ as the cluster head. All the neighbors of the chosen cluster head are no longer allowed to participate in the election procedure.

Step 8. Repeat steps 2---7 for the remaining nodes not yet selected as a cluster head or assigned to a cluster.

### IV. PROPOSED WORK

*Factors that influence the implementing the GA*

A brief discussion of four factors is given below:

1. *degree-difference*: $\Delta_v = |d_v - \delta|$ for every node $v$. Here δ is ideal node number of a cluster except the cluster head.

2. Battery power ($P_v$): Obviously, the higher the battery power, the higher the probability that the node will become CH.
3. Degree of mobility: The mobility of the node has great impact on the network lifetime. The topology of the network will be change very frequently due to the high mobility of nodes, which leads to reselection of CHs rapidly.
4. *sum of the distances*, $D v$ with all its neighbors, as

$$D_v = \sum_{v' \in N(v)} \{dist(v,v')\}$$

*Optimization Approach For Cluster Head Selection Using GA:*

**Algorithm:**
```
Alg. Clustering_GA(int chromosome[][] )
 {
   Take dataset(chromosome matrix) according to the node's
neighbourhood at time t;
while(not end of all chromosome in chromosome matrix)
   {
    Take the first row(chromosome) from chromosome matrix;
    Generate the Gene matrix using the parameter Δv, Dv, Mv
   Pv from the first chromosome row;
    while(convergence criteria is not met )
   {
    Calculate the Wv , value for each Gene (For i=1 to 4)
    {  Wvi = w1Δv + w2Dv + w3Mv + w4Pv
       Wv , = Wv + Wvi
       If(i==4)
        { j=1;
          b[j]= Wv
          j++;
        }
    }
      Maximum and Minimum value is taken from b array;
      Minimum value of b array position row is replaced
      Maximum value of b array  position row;
      Getting a new Gene matrix ;
      Take two parent from Gene matrix;
      Mod_Gene[][]=Crossover(Gene);
      Mutation(Mod_Gene[][]);
     }/End For/
    }/End While/
  One of the CH is choosen from the chromosome;
   Take another chromosome;
  }/End main while/
  A set of CH will be choosen among the data set;
  The duplicate node in the set will be deleted to get the
  desired result;
 }/End of alg./
```

## III. METHODOLOGY

Our goal is to search best nodes among hundreds of nodes, so that they can act as CHs.
Conventional search methods are not robust, while the GA is a search procedure that uses random choice as a tool to guide a highly exploitative search through a coding of a parameter space. According to Goldberg the GA has 4 major characteristics:
1. GAs with a coding of the parameter set, not the parameters themselves.
2. GAs search from a population of points, not a single point.
3. GAs use payoff (objective function) information, not derivatives or other auxiliary knowledge.
4. GAs use probabilistic transition rules, not deterministic rules.

In many optimization methods, we move carefully from a single point in the decision space to the next using some transition rule to determine the next point. This point-to-point method is dangerous because it is a perfect prescription for locating false peaks in multi modal (many peaked) search spaces. By contrast, GA works from a rich database of points simultaneously (a population of strings), climbing many peaks in parallel; thus, the probability of finding a false peak is reduced. A GA starts with a population of strings and thereafter generates successive populations of strings. A simple GA consists of three operators:
1. Reproduction
2. Crossover
3. Mutation

The chromosome of the GA contains all the building blocks to a solution of the problem at hand in a form(fig-1) that is suitable for the genetic operators and the fitness function. Each individual node is represented by a 4 number called `gene'. These four parameter which define the feature of the node and are represented as follows:

Node ID → X1   X2   X3   X4

X1: degree-difference
X2: Battery power (Pv),
X3: its degree of mobility, and
X4: sum of the distances

Let's take an example. To start off, select an initial chromosome of total population are neighbours of particular node ID . Here, we select a population of size equal to the no of nodes . Then we have to operate on each chromosome using the 4 parameter for each neighbor nodes of particular node ID. Corresponding node ID has a cluster haead that sould be determined by some fitness value. This value can be evaluated from a fitness function,

$f(x) = f(x1; x2; x3; x4) = W_1*v + W_2*P_v + W_3*M_v + W_4*D_v.$

case of Ad-hoc the fitness function depends upon the four factors, discussed in above. And minimum of $f(x)$ should be selected as cluster head. A generation of the GA begins with reproduction. We select the mating pool of the next generation by spinning the weighted roulette wheel four times. From this, the best string get more copies, the average stay even, and the worst die off. Above procedure should be applied for each of the chromosome.

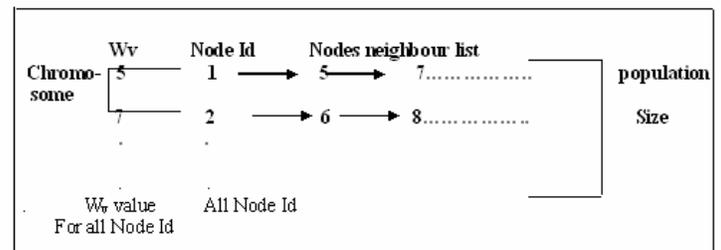

Figure(4.1)( Array Representation of Nodes)

## IV. GRAPHICAL ANALYSIS

Here, we have shown the comparison between deterministic approach and GA-based approach of weighted clustering algorithm. And we see that sometime genetic algorithm based approach is better than the deterministic approach which is shown in figure( 6.5).and sometime show both approach produces the same number of clusterheads as well as cluster. Sometime deterministic gives the lower number of cluster than the number of cluster in GA-based approach. In figure(6.5) green color curve represents the deterministic approach of clustering and yellow color curve represents the GA-based approach .How average number of cluster are changing with respect to the varying transmission range with fixed displacement equal to 5

In figure (6.6) shows the comparison of deterministic and GA-based approach between average number of cluster and varying displacement. and we see that GA-based approach always provides the better result than the deterministic approach.

In figure(6.7) shows the comparison of deterministic and GA-based approach between Connectivity and Transmission range .Here connectivity can be measured by the probability that a node is reachable to any other node. For a single component graph ,any node is reachable to the any other node and the connectivity is 1.If the network does not not result in single component graph, then we can say that all the other node in the largest component can communicate with each other and the connectivity can be ratio of the cardinality of the largest component to the cardinality of the graph. From figure(6.7) we have shown the transmission range of the cluster head can be large enough to yield the connected network. If we compare the deterministic approach and GA-based approach ,there we have shown GA gives the better connectivity than the deterministic approach. A well connected graph can be obtained at the cost of a higher transmission range. If we see the graph of transmission range versus average number of cluster heads. There we can see the cluster head will be minimum by incrementing the transmission range .But in GA-based approach gives the better result than deterministic approach. So that in respect of connectivity ,GA-based approach gives the better result.

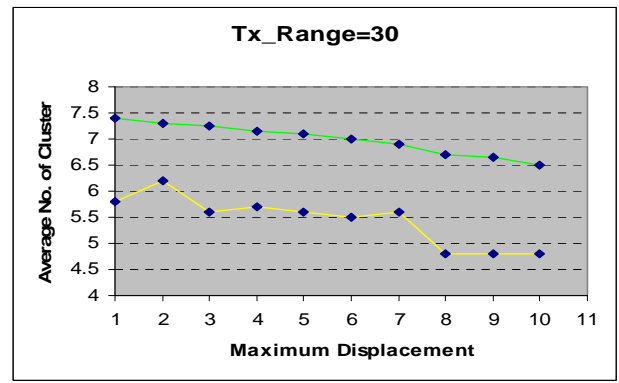

Green Curve = Deterministic Approach.
Yellow Curve = GA-based Approach

**Figure(6.6)Comparison Between Deterministic and Soft Computing Approach With Fixed Transmission Range)**

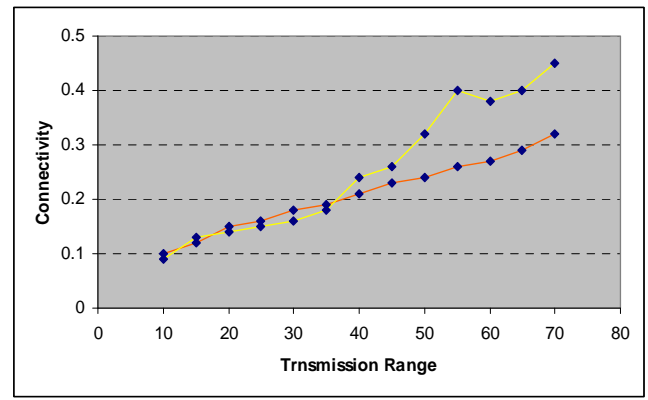

Yellow Color Curve= GA-Based Approach
Red Curve = Deterrministic Approach
**Figure(6.7) Connectivity Vs Transmission Range**

## V. CONCLUSION

From the graphical analysis, we have done comparison analysis between deterministic WCA and GA-based WCA and there we have seen that, we can not get always optimistic result in genetic algorithm because genetic algorithm is a randomized searching technique. We have seen when transmission range increases then average number of clusters decreases (Figure(6.5)),so that connectivity of network should be better to compare with the deterministic WCA.

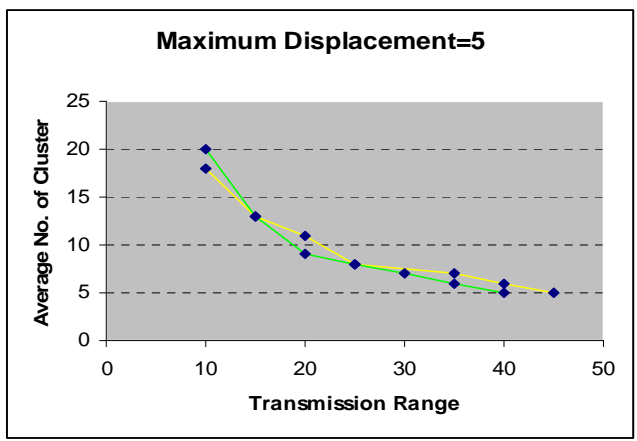

Green Curve = Deterministic Approach
Yellow Curve = GA-based Approach

**Figure(6.5)(Comparison Between Deterministic and Soft Computing Approach with Fixed Displacement)**

VII. AUTHORS' PROFILE

**Bhaskar Nandi** is a lecturer with the department of Computer Science and Engineering, Seacom Engineering College,Howrah,Kolkata,West Bengal, India. He has a teaching experience of about two years, and 1 year of research experience, ,more than two years of industry experience. His research interest are in soft computing, Ad-hoc Networking, Information Security and Data Mining. He has publication in different national journal and conferences. Presently he is working Data Mining and Network Security.

**Subhabrata Barman** is a Senior lecturer with the department of Computer Science and Engineering, Haldia Institute of Technology, Haldia, West Bengal, India. He has a teaching experience of more than 6 years and a research experience of more than 2 years. His research interests are in the field of Mobile Networking and Computing, Computational Intelligence, Image Processing, Speech and Signal Processing. He has several publications in several national and international conferences and journals. Currently he is working in the area QoS issues and Energy Management in Wireless Adhoc and Sensor Networks.

**Soumen Paul** is an Assistant Professor with department of the Information Technology, Haldia Institute of Technology, Haldia, West Bengal, India. He has a teaching experience of more than 8 years, industry experience of 11 years and a research experience of more than 2 years. His research interests are in the field of Control Engineering, Soft Computing and Mobile Networking. He has publications in several national and international conferences and journals. His doctoral work is in the area of Deadbeat realization of linear, non-linear, time invariant control systems of nth order.